\documentclass[%
 reprint,
 amsmath,amssymb,
 aps,
prl,
]{revtex4-1}

\usepackage{color}
\usepackage{graphicx}
\usepackage{dcolumn}
\usepackage{bm}
\usepackage{amsmath}
\usepackage[pdftex,plainpages=false,colorlinks=true,linkcolor=blue, citecolor=blue, urlcolor=blue]{hyperref}

\newcommand{\dx}{{d_{x^2-y^2}}}
\newcommand{\dz}{{d_{z^2}}}

\newcommand{\lno}{{\mathrm{La_3Ni_2O_7}}}

\begin{document}

\preprint{APS/123-QED}

\title{Superconductivity in the Bilayer Two-orbital Hubbard Model
}

\author{Yao-Yuan Zheng}

\affiliation{
 School of Physics, Sun Yat-sen University, Guangzhou, Guangdong 510275, China
}

\author{W\'ei W\'u}
\email[Corresponding author: ]{wuwei69@mail.sysu.edu.cn}
\affiliation{ School of Physics, Sun Yat-sen University, Guangzhou, Guangdong 510275, China
}

\date{\today}


\begin{abstract}
Motivated by the recently discovered high $T_c$  nickelate superconductor $\mathrm{La_3Ni_2O_7}$, we investigate superconductivity in a two dimensional Hubbard model on square lattice that consists of two layers of hybridizing $d_{x^2-y^2}$ and $d_{z^2}$ orbitals. Employing cluster dynamical mean-field theory, we establish phase diagrams resolving the  crucial dependence of superconducting $T_c$ on electron hybridization  $V$ between $d_{x^2-y^2}$ and $d_{z^2}$ orbitals at the same layer, and on hopping $t_{\perp}$ between two $d_{z^2}$ orbitals at different layers.  $V$  and  $t_{\perp}$ are presumably  linked to the pairing phase coherence and the ``pairing glue'' of the system respectively. Our result in general favors a two-component theory explanation of superconductivity in a composite system. The influence of the pseudogap effect and  Hund's coupling on superconductivity are discussed, and also
the implication of our result to the understanding of $\mathrm{La_3Ni_2O_7}$ superconductivity.
\end{abstract}

\keywords{Nickelates superconductivity, Hubbard model, Dynamical mean-field theory, quantum Monte Carlo}

\maketitle

\textit{Introduction - } Since the discovery of heavy fermion and cuprate superconductors~\cite{steglich1979, bednorz1986}, unraveling unconventional superconductivity (SC)~\cite{ sun2023, zhang1988,patrick06, haule2007,  Scalapino_RMP12,chubu2008, keimer2015quantum,htsc21_nrp,kowalski2021oxygen} has been one of the thematic topics in condensed matter physics. 
For cuprate, despite its precise nature of $d-$wave pairing remains a mystery, it is widely believed that 
its essential physics may be captured by the single-band Hubbard model or the $t-J$ model encoding the correlation effects of the Cu-$3d_{x^2-y^2}$ electrons~\cite{haule2007,Scalapino_RMP12}.
The sinlge-band Hubbard model also captures many other unusual phenomena of correlated electrons~\cite{patrick06,haule2007,Scalapino_RMP12,zheng17,wu2022,Raghu2022}, which makes it one of the most intenstively studied theoretical models in condensed matter physics.
It is worth noting that, many other superconductors, including the  layered organic superconductor~\cite{mckenzie1997} and the  transition metal dichalcogenides (TMDs)~\cite{tmd18}, can also be described by the single-band Hubbard model.

\begin{figure}[t!]
\includegraphics[scale=1.1]{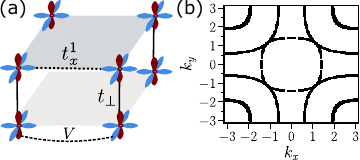}

\caption{Geometry and Fermi surface of the bilayer two-orbital Hubbard model on two dimensional square lattice. In subplot \textbf{(a)} the light symbols denote $\dx$ orbitals while dark symbols denote $\dz$ orbitals.  To plot the Fermi surface in subplot \textbf{(b)} , $t^{1}_x =-1, t^{2}_x = 0.15, t^{1}_z =-0.25, t^{2}_z = -0.05, \mu_x = -1.6, \mu_z = -0.8, t_{\perp} =1.3, V= 0.5 $ are used.}
\label{fig:illu}
\end{figure}

Very recently, novel high $T_c$  superconductivity is reported in the pressurized  nickelate $\mathrm{La_3Ni_2O_7}$\cite{sun2023,liu2023,zhang2023high,hou2023emergence,li2023structural,li2023signature,cui2023strain,sun2023evidence}, which has a layered structure consisting of double $\mathrm{NiO_2}$ planes. As revealed by the density functional theory (DFT) calculations~\cite{pardo2011,luo2023bilayer,lechermann2023electronic,zhang2023trends,geisler2023structural,Werner23} and experiments~\cite{liu2023electronic}, the $e_g$ doublet of  Ni-$3d$ orbitals, \textit{i.e.}, the $3d_{x^2-y^2}$ and $3d_{z^2}$ orbitals, are both at play 
 in low-energy physics.  The $d_{z^2}$ orbital can be seen as more localized and near half-filled, whereas
 the $d_{x^2-y^2}$ orbital is near quarter-filling and in general, more itinerant~\cite{yang2023orbital,wu2023charge}. Electrons at two different $e_g$ orbitals can hybridize via an intra-layer hopping $V$ between them. The  out-of-plane electron hopping $t_{\perp}$ between $d_{z^2}$ orbitals at different layers, generates an effective antiferromagnetic (AFM) coupling $J_{\perp}$  deemed to be the major source of the correlation effects in the system (see Fig.~\ref{fig:illu})~\cite{zhang2023eff,wu2023charge,ryee2023critical}. The above picture can be in general summarized as a bilayer two-orbital model~\cite{luo2023bilayer} of  $\mathrm{La_3Ni_2O_7}$, which 
  differs essentially from the Hubbard model of cuprates, or that of the infinite layer nickelates $R\mathrm{NiO_2}$ ($R$=La,Pr,Ni)~\cite{Lidanfeng_nature19} which to be captured by the $d_{x^2-y^2}$ orbital hybridizing with less correlated $5d-$derived bands~\cite{wang2020dist,karp20,nomura2022,kitatani23}.
  It is also argued that the Hund's coupling $J_H$ may play an important role in $\lno$~\cite{oh2023type,lu2023interlayer,liao2023electron}.
  All in all, the low-energy physics of $\mathrm{La_3Ni_2O_7}$ can be sharply distinguished from previously known unconventional superconductors, which goes beyond the description of the single-band Hubbard model. A systematic study of SC in the bilayer two-orbital Hubbard  (BLTO-Hubbard) model will not only shed light on comprehending  $\mathrm{La_3Ni_2O_7}$ SC, but aslo provide new insights into the general understanding of unconventional superconductivity.
 
In this work, we investigate superconductivity in BLTO-Hubbard model using cluster extensions of the dynamical mean-field theory (DMFT)~\cite{georges96,maier05rmp}. 
Performing finite temperature computations, we reveal following key aspects of SC in this system: 
(1) The $s_{\pm}-$wave pairing~\cite{yang2023possible, tian2023correlation,gu2023effective,lu2023interplay,qu2023bilayer,qin23,luo2023RMFT,sakakibara2023possible,pan2023effect,jiang2023high,chen2023orbital} driven by the inter-layer AFM coupling between $d_{z^2}$ orbitals with effective  $J_{\perp} \propto t_{\perp}^2/U_{z}$,  has a non-monotonic dependence on $t_{\perp}$. In small $t_{\perp}$ regime, superconducting $T_c$ grows with $t_{\perp}$ (or $J_{\perp}$), whereas at large $t_{\perp}$, $T_c$ can be suppressed by increasing  $t_{\perp}$ ( or $J_{\perp}$). Here $U_z$ is the Hubbard repulsion within $\dz$ orbital.
(2) There exists a critical $V_{c}$ only when the $\dz$-$\dx$ hybridization $V$ greater than which the system can be superconducting. The relation between parameter $V$ and phase coherence of pairing is discussed.  (3) The calculated superconducting $T_c$ for parameter set in relevance to pressurized $\mathrm{La_3Ni_2O_7}$ in general agrees with the experimental $T_c$~\cite{sun2023}. Increasing $V$ leads to higher superconducting $T_c$.
The connection between our result and  the two-component theory of superconductivity in a composite system~\cite{yang23inter,kivelson2002making,berg08}  is  discussed.

\textit{Model and method - } The bilayer two-orbital Hubbard model~\cite{luo2023bilayer} we use can be written as $H = H_0 + H_U$, which reads,
\begin{eqnarray}
H_0 =-t_{x}^{ij}\sum_{ ij \alpha\sigma}c_{i\alpha\sigma}^{\dagger}c_{j\alpha\sigma}-t_{z}^{ij}\sum_{ ij\alpha\sigma}d_{i\alpha\sigma}^{\dagger}d_{j\alpha\sigma} \nonumber \\ 
-t_{\perp}\sum_{i\sigma \alpha}d_{i\alpha\sigma}^{\dagger}d_{i\overline{\alpha}\sigma} 
+V\sum_{\langle ij\rangle\alpha\sigma}c_{i\alpha\sigma}^{\dagger}d_{j\alpha\sigma} \nonumber  \\
- \mu_x\sum_{i,\alpha,\sigma}n_{i\alpha\sigma}^{x} - \mu_z\sum_{i,\alpha,\sigma}n_{i\alpha\sigma}^{z} \\
H_U= U_{z}\sum_{i\alpha}n_{i\alpha\uparrow}^{z}n_{i\alpha\downarrow}^{z}
\end{eqnarray}

where $t_{x}^{ij}$, $t_{z}^{ij}$ are the in-plane (or intra-layer) intra-orbital  electron hoppings for $\dx$ and $\dz$ orbitals respectively. $\alpha$ enumerates the two layers ($\alpha=1,2$) and  $\sigma$ indicates spin up/down.  We set
the in-plane hopping between the nearest neighbor (NN) sites of $\dx$ orbitals $t_{x}^{1} \equiv t =1$ as the energy unit throughout the paper, $t_{x}^{1} \sim 0.48eV$ according to DFT downfolding~\cite{luo2023bilayer}. Following which~\cite{luo2023bilayer} the in-plane next-nearest neighbor (NNN) hopping of $\dx$ orbitals is taken as $t_{x}^{2} = 0.15t$. The in-plane NN hoping between $\dz$ orbitals $t_{z}^{1}= -0.25$, and NNN hopping $t_{z}^{2}= -0.05$.
The orbital energies $\mu_x$ and $\mu_z$ can be changed to adjust the filling factors of the two 
$e_g$ orbitals~\cite{wu2023charge}, while the total filling $n = \langle n^z_{i\sigma}+n^x_{i \sigma} \rangle = 0.75$ is fixed. The hybridization between $\dz$ and $\dx$ orbital at the same layer $V$, as 
well as the  hopping  $t_{\perp}$ between two  $\dz$ orbitals at the same site but different layers, will be the main controlling parameters in our study (see also Fig.\ref{fig:illu}).

For the interacting part $H_U$, we will first
consider only the Hubbard repulsion $U_{z} =7$ between the $\dz$ electrons,  the more itinerant $\dx$ orbital is to be taken as non-interacting. In the end of the work, the Hubbard repulsion within $\dx$ orbital $U_x$, inter-orbital Hubbard $U_{xz}$, as well as a longitudinal Hund's coupling $J_H$ will be considered. 

The cluster DMFT methods have been extensively used in studying cuprates and related problems for they can effectively capture temporal and short-ranged spatial correlations~\cite{maier05rmp,staar14}. Depending on the boundary conditions of the effective cluster, there can be dynamical cluster approximation (DCA) and cellular DMFT (CDMFT) variations~\cite{maier05rmp}. 
As we will show below, our results using different clusters converge rapidly against cluster size, demonstrating unambiguously the solidness of the cluster DMFT methods on computing $T_c$ in this system.
We will use the continuous time quantum Monte Carlo (CTQMC)~\cite{rubtsov05} as impurity solver. For cases with severe sign problem, as will be explicitly stated, Hirsch-Fye quantum Monte Carlo (HFQMC) impurity solver is also used. We have carefully verified that 
the two impurity solvers give the same result. Determinant quantum Monte Carlo (DQMC) is also employed to benchmark our cluster DMFT result at high temperatures.

\begin{figure}[t]
\includegraphics[scale=0.55]{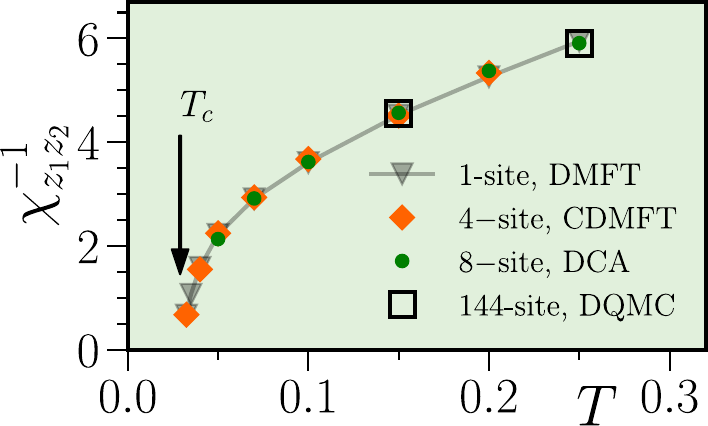}

\caption{Inverse of the inter-layer $\dz - \dz$  component of the $s_{\pm}$-wave pairing susceptibility, $[\chi_{z1,z2}]^{-1}$ as a function of temperature $T$. The temperature where $[\chi_{z1,z2}]^{-1} \rightarrow 0$ is defined as the superconducting transition $T_c$ (black arrow). Here $U_z = 7, n_z = 0.47, n_x=0.28, t_{\perp} = 0.8, V=0.5$. HFQMC impurity solver is used here. }
\label{fig:chi0}
\end{figure}

\textit{Pairing susceptibility-} For a multi-band system, different components of the pairing susceptibility
 simultaneously diverge as approaching superconducting $T_c$~\cite{wu15}. Without loss of generality, here we select a $\dz$-$\dz$ component of the pairing susceptibility, $\chi_{z_{1}z_{2}}$, in which the dominant branch of the $s_{\pm}$ wave pairing is found (see discussion below), to identify the pairing $T_c$,

\begin{equation}
\chi_{z_{1}z_{2}}=\frac{1}{N}\sum_{i,j}\int_{0}^{\beta}\langle\mathcal{T_{\tau}}p_{j}^{\dagger}(\tau)p_{i}(0)\rangle d\tau
\end{equation}

where $p_i$ is the on-site inter-layer spin-singlet pairing operator $p_i = 1/\sqrt{2}(d_{i,1,\uparrow}d_{i,2,\downarrow}-d_{i,1,\downarrow}d_{i,2,\uparrow}) $ of the $\dz$ orbital. Fig.~\ref{fig:chi0} plots the 
inverse pairing susceptibility  $\chi^{-1}_{z_{1}z_{2}}$ as a function of temperature $T$ for $\dz$, $\dx$ orbtral  being respectively close to half-, and quarter-filled ($n_z \sim 0.5$, $n_x \sim 0.25$). One sees that
as lower temperature $T$, $\chi^{-1}_{z_{1}z_{2}}$ decreases and eventually extrapolates to zero as approaching $T_c \sim  0.029$  (black arrow), indicating that the spontaneous SC instability can happen in this system. In Fig.~\ref{fig:chi0}, we find that different cluster DMFT schemes, including the single-site ($1\times 4$ orbitals) DMFT, four-site ($4\times 4$ orbitals) CDMFT, and eight-site ($8\times 4$ orbitals) DCA, all give results in 
excellent agreement with one another, which are agreed by the $144$ site ($12 \times 12 \times 4$ orbitals ) DQMC simulations at $T =0.25,0.15$. This result demonstrates that the non-local fluctuations are insignificant in the BLTO-Hubbard model in the given parameter regime, and also indicates the solidness of the cluster DMFT methods on computating $T_c$~\cite{staar14} here. 
This observation contrasts the cluster DMFT results of the single-band Hubbard model, where finite-site effects are typically strong in small  clusters~\cite{staar14}. 

\begin{table}[b]
\begin{tabular}{ccccc}
\hline \hline 

  & $\langle d^{\dagger}_{i, z_1 \uparrow} d^{\dagger}_{i , z_2 \downarrow} \rangle$ \quad  & $\langle c^{\dagger}_{i,x_1 \uparrow} c^{\dagger}_{i, x_2 \downarrow} \rangle$ \quad &$\langle d^{\dagger}_{i,z_1 \uparrow} d^{\dagger}_{i+1, z_1 \downarrow} \rangle$  & $\langle c^{\dagger}_{i,x_1 \uparrow} c^{\dagger}_{i+1, x_1 \downarrow} \rangle$ \tabularnewline\hline 

 &  0.011  & 0.014  & 0.002 & 0.0004\tabularnewline  \hline \hline

\end{tabular}
\caption{\label{tab:hop1} Different components of the superconducting order parameter in real-space at $U_z = 7, n_z = 0.47, n_x = 0.28,t_{\perp} = 0.8, V=0.5$ (same as in Fig.~\ref{fig:chi0}) from $4-$site ($2\times 2$) CDMFT,  $T = 0.025 < T_c \approx 0.029$.  Here $\langle d^{\dagger}_{i, z_1 \uparrow} d^{\dagger}_{i , z_2 \downarrow} \rangle$ is the on-site, inter-layer, $\dz$ orbital component, while $\langle c^{\dagger}_{i,x_1 \uparrow} c^{\dagger}_{i+1, x_1 \downarrow} \rangle$ denotes the intra-layer $\dx$ orbital component  on  nearest neighbor bond. HFQMC solver is used here. } 
\label{tab:1}
\end{table}

Further decreasing  temperature $T$,  one obtains finite SC order parameter $\Delta^{\alpha,\beta}_{ij} = \langle \phi^{\dagger}_{i \alpha \uparrow} \phi^{\dagger}_{j \beta \downarrow} \rangle$, $\phi^{\dagger}_{i \alpha \sigma} \in \{ c^{\dagger}_{i \alpha \sigma}, d^{\dagger}_{i \alpha \sigma} \}$ in real space when $T < T_c$. As shown in Table~\ref{tab:1}, we find that the local inter-layer ($R_i = R_j, \alpha \neq \beta$ ) components overweight all other non-local ($R_i \neq R_j$) branches,  suggesting that the SC can be best described as an $s_{\pm}-$ wave pairing here, which agrees with previous studies~\cite{yang2023possible, tian2023correlation,lu2023interplay,qu2023bilayer,qin23,luo2023RMFT,qu2023roles}.

\begin{figure}[t]
\includegraphics[scale=0.6]{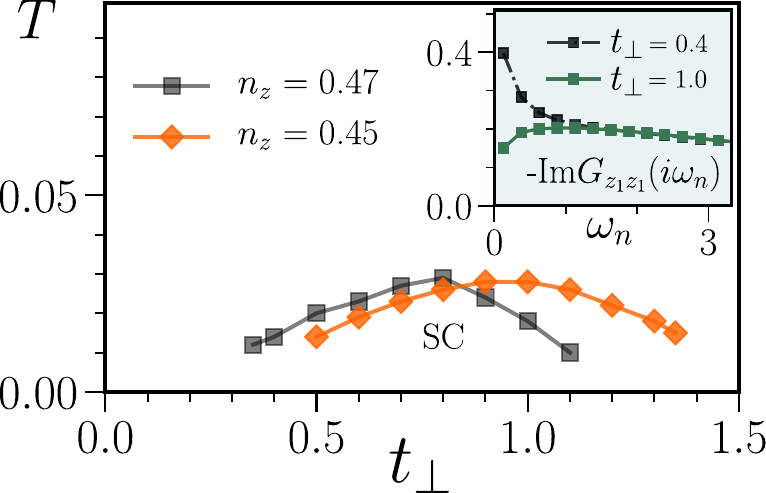}

\caption{ Superconducting transition temperature $T_c$ as a function of  the vertical hopping amplitude $t_{\perp}$ between two inter-layer $d_{z^{2}}$ orbitals.Results for two different dopings are shown. \textbf{Inset:} Imaginary part of the local Green's function of $\dz$ orbital $\mathrm{-Im}G_{z_1z_1}(i\omega_n)$ are shown as a function of Matsubara frequency $\omega_n$  for $n_z =0.47$ at two different $t_{\perp}$. Here $V = 0.5, U_{z} = 7, n_x = 0.75-n_z$.} 
\label{fig:phasev}
\end{figure}

\textit{Phase diagram-} We now study the dependence of $T_c$ on the inter-layer $\dz$-$\dz$ hopping $t_{\perp}$. In the localized and large $U_z$ limit of $\dz$ orbital, the effective AFM exchange $J_{\perp}$ between the two $\dz$ orbitals scales like $J_{\perp} \sim t_{\perp}^2/U_z$  to dominant order. As found in cuprates~\cite{ruan2016relationship,kowalski2021oxygen}, one may expect that the superconducting $T_c$ grow with magnetic coupling $J_{\perp}$, or equivalently with $t_{\perp}$ here. This is indeed what we observe  in the small $t_{\perp}$ regime in Fig.~\ref{fig:phasev}, where $T_c$ for two different dopings,  $n_z=0.47$ (squares) and $n_z=0.45$ (diamonds) are shown versus $t_{\perp}$. At large $t_{\perp}$, saying $t_{\perp} \gtrsim 0.8 $ for $n_z=0.47$, we however find that $T_c$ decreases as  increasing $t_{\perp}$. In other words, non-monotonic  dependence of $T_c$ on $t_{\perp}$ (or $J_{\perp}$ ) is observed in Fig.~\ref{fig:phasev}. This behavior appears to resemble the superconducting dome in the BCS-BEC crossover phase diagram~\cite{chen2005bcs}. To gain insight into this observation,  Inset of  Fig.~\ref{fig:phasev} compares the imaginary part of the local Green's function of the $\dz$ orbital $\mathrm{-Im}G_{z_1,z_1}(i\omega_n)$
for two different $t_{\perp}$ at $n_z=0.47$. As one can see that at large $t_{\perp}$ ($t_{\perp}=1.0$), the low-energy weight of the Green's function in the normal state ($T=0.04 > T_c$) is greatly suppressed comparing to the smaller $t_{\perp} = 0.4$ case. Further analysis finds that  the suppression of Green's function in the normal state at large $t_{\perp}$ corresponds to the opening of a pseudogap in the low-energy density of states (not shown here).  Verifying whether 
the BCS-BEC crossover theory applies in this model requires identifying the physical origin of the pseudogap, which can be done, for example, by using the fluctuation diagnostics method~\cite{Gunnarsson15,wu2022} in future studies. It is worth noting that in cuprate, it is generally believed that the SC state is far from  the BEC limit~\cite{sous2023absence}, despite the existence of pseudogap.
Summarizing Fig.~\ref{fig:phasev}, we find the tendency that as $n_z$ decreases, the suppression 
of the low-energy local Green's functions at large $t_{\perp}$ becomes less severe, and the dome-like superconducting regime as a whole moves to the large $t_{\perp}$ side. 


\begin{figure}[b]
\includegraphics[scale=0.5]{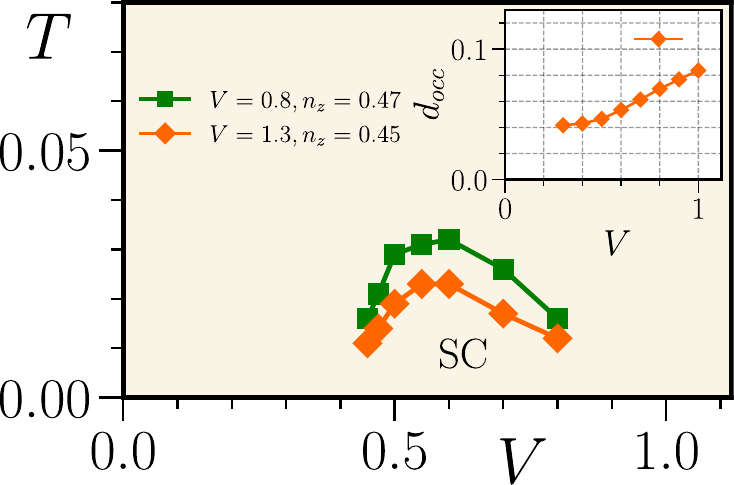}

\caption{ Superconducting $T_c$ as a function of $V$ are shown for two different sets of ($t_{\perp}$,$n_z$) parameters (see also Fig.~\ref{fig:phasev}). \textbf{Inset:} The double occupancy $d_{occ} = \langle n^{d}_{\uparrow} n^{d}_{\downarrow}  \rangle$  of $\dz$ orbital as a function of $V$ at ($t_{\perp}=1.3$,$n_z=0.45$).   $d_{occ}$  increasing with $V$ suggests the growing itinerancy of $\dz$ orbital as  $V$ increases. Here $U_z = 7, n_x=0.75-n_z$.
 } 
\label{fig:txz}
\end{figure}

Now we focus on how the $\dz$-$\dx$ hybridization $V$ impacts on superconductivity. 
It is obvious that the SC instability found above cannot
be attributed to the $\dz$ orbital exclusively, though the binding of electrons does stem from the effective $J_{\perp}$ in this orbital.
Indeed, here the $\dz$ orbital has a small intra-orbital NN hopping $t^{1}_{z}/t = -0.25$, the Cooper pairs face difficulty in propagating coherently within the  $\dz$ plane to achieve phase ordering. The more itinerant  $\dx$ orbital, on the other hand, can boost phase coherence of Cooper pairs via $\dz$-$\dx$ hybridization $V$~\cite{yang23inter,berg08}.  Fig.\ref{fig:txz} displays superconducting $T_c$ as a function of $V$  for two different  ($t_{\perp}$,  $n_z$) parameter sets. It is clear in Fig.\ref{fig:txz} that  finite $T_c$
can be obtained only when $V$ is sufficiently large. Extrapolating $T_c$ curve to zero $T$, taking the ($t_{\perp}=0.8, n_z = 0.47$ ) case for example, a critical  $V_{c} \sim 0.4$ can be inferred for SC instability in the zero-$T$ limit. As for $V > V_{c}$, $T_c$ grows rapidly  with $V$, which reaches a maximum $T_c \sim 0.03$  for the $t_{\perp}=0.8$ case. Further increasing $V$ leads to, nevertheless, a suppression of $T_c$. This is because
a large  $V$  can delocalize $\dz$ electrons, resulting in the disruption of  local   $\dz$ spin singlets and thereby being detrimental to SC. The growing itinerancy of $\dz$ electrons with $V$ can be  manifested  in the Inset of Fig.\ref{fig:txz}, where double occupancy $d_{occ}$ of $\dz$ orbital is shown as a function of $V$.  In Fig.~\ref{fig:txz}, a different parameter set of ($t_{\perp}=1.3, n_z =0.45$) is also shown(diamonds), which exhibits similar $V-$ dependence  of $T_c$, suggesting the generality of our above analysis on the role of $V$.

It has been proposed by
 Yang and collaborators~\cite{yang23inter,qin23} that  SC in the  pressurized $\lno$ fits in a two-component theory. That theory was first established by Kivelson~\cite{kivelson2002making,berg08} to describe a ``composite system'' of superconductivity which can be divided into two distinct components: a ``pairing'' component with high pairing energy scale $\Delta_0$ but vanishing superfluid stiffness (such as the $\dz$ orbital here), and a ``metallic'' component with no pairing but high superfluid stiffness ( like the $\dx$ orbital here).  Berg \textit{et al}~\cite{berg08} show  that a high  $T_c $ can be reached approaching its mean-field value, $T_c \sim T_{MF} \approx \Delta_0/2$ if a moderate hybridization between the two components is introduced to suppress the phase fluctuations. In Fig.~\ref{fig:chi0}, we  indeed see that  in the BLTO-Hubbard model the non-local fluctuations beyond the single-site DMFT  appear to be unimportant.  It is also interesting to notice that here the optimal $T_c$ is found near hybridization $V/t \approx (0.55 \sim 0.60)$ for the two parameter sets we study, which coincides with the value $V/t \sim 0.5$ obtained for the attractive Hubbard model with $U/t=-1$~\cite{berg08}. One prominent difference is that, however, we find a finite critical $V_c \approx 0.4$ for SC, while in Ref.~\cite{berg08} SC vanishes
 continuously in the $V \rightarrow 0 $ limit. This may be attributed to
 the temporal fluctuations which are treated exactly in CDMFT here, but absent in the work of Ref.~\cite{berg08}.

\begin{figure}
\includegraphics[scale=0.5]{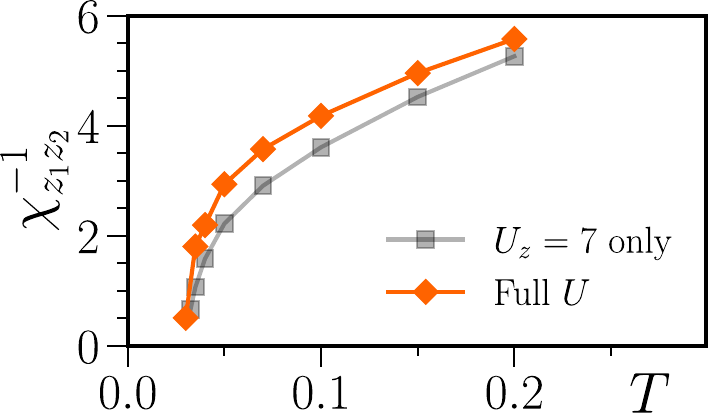}

\caption{Inverse pairing susceptibility, $[\chi_{z1,z2}]^{-1}$ as a function of temperature $T$ for two different configurations of Hubbard parameters. ``Full $U$'' denotes the case with  $H^{\prime}_U$ and $U_z$ both considered, \textit{i.e.}, $U_x = U_z =  7$, $U_{xz} = U_z - 2J_H$, $J_H = 1.4$. Single-site DMFT and HFQMC impurity solver are used here.}
\label{fig:chi1}
\end{figure}
 
\begin{figure}[t]
\includegraphics[scale=0.42]{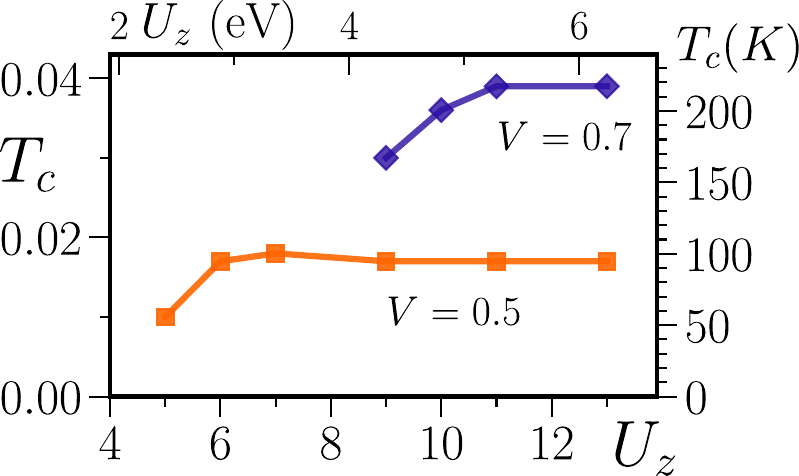}

\caption{ Superconducting $T_c$ as a function of  $U_{z}$ for two different $V$ at fixed $ t_{\perp}=1.3, n_z=0.45, n_x=0.3$. 
The energy unit $t^{1}_{x}$ is set as $t^{1}_{x} = 0.48 \mathrm{eV} $ as following the DFT downfolding~\cite{luo2023bilayer}. 
} 
\label{fig:U}
\end{figure}

\textit{Effect of $U_{x}$, $U_{xz}$ and $J_H$ - } Above we have discussed SC in the BLTO-Hubbard model considering only $U_{z}$~\cite{zhang2023eff}. To draw a closer connection to the realistic materials, other Hubbard parameters, including Hubbard repulsion $U_{x}$ within the $\dx$ orbital, $U_{xz}$ between the $\dz$ and $\dx$ orbital on the same site, as well as a longitudinal Hund's coupling $J_H$ are considered below as, $H^{\prime}_U= U_{x}\sum_{i\alpha}n_{i\alpha\uparrow}^{x}n_{i\alpha\downarrow}^{x}
+ U_{xz}\sum_{i\alpha}n_{i\alpha \sigma }^{x}n_{i\alpha \overline{\sigma} }^{z} 
+(U_{xz}-J_H)\sum_{i\alpha, \sigma }n_{i\alpha\sigma}^{x}n_{i\alpha \sigma}^{z}$. In Fig.~\ref{fig:U}, the inverse pairing susceptibility $\chi_{z1,z2}^{-1}$ as a function of $T$ is shown to compare the results with $H^{\prime}_U$  included(diamonds), and without  $H^{\prime}_U$ (only $U_z$, squares). As shown in which, we find that including $H^{\prime}_U$ ( $U_{x}, U_{xz} $ and  $J_H$ ) in our study in general only modifies the $T_c$ in a quantitative way. The superconducting state is not significiantly changed.
We note that a similar conclusion on the role of $J_H$  is made by the renormalized mean-field theory study on SC in the bilayer two-orbital $t-J$ model~\cite{luo2023RMFT}. 
A recent numerical study~\cite{wu2023charge} on the more $\textit{ab initio}$ Emery-Hubbard model also finds that Hund's coupling $J_H$ does not affect much the magnetic correlations of the system due to the small filling factor of $\dx$ orbital.

\textit{Connection to $\lno$ and raising $T_c$-} Finally, we discuss the connection between our result and the $\lno$ materials.  
 For pressurized $\lno$, DFT result suggests $t_{\perp} \approx 1.3, V \approx 0.5$ for the BLTO-Hubbard model~\cite{luo2023bilayer}. However, the Hubbard parameters such as  $U_z$  have not been reliably estimated. In  Fig.~\ref{fig:U} we therefore plot $T_c$ as a function $U_z$ while fixing $t_{\perp} =1.3, V = 0.5$ (squares). One can see that as $U_{z} > 6$, $T_c$ is almost independent of $U_z$ in the $U_z$ range we consider, suggesting a $T_c \approx 100$ Kelvin. This result in general agrees  with the experimental  $T_c \sim 80 K$. Our study is also consistent with the absence of SC in  $\lno$ under ambient pressure where $\dz$ orbital is below the Fermi level~\cite{yang2023orbital}:  the $\dz$ orbital is the driving force of the composite SC. The pairing will be disabled if $\dz$ orbital is not active at low-energies. Finally, Fig~\ref{fig:U} shows that $T_c$ can be considerably enhanced by increasing $V$ if $U_z$ is sufficiently large ($U_z \gtrsim 4 \mathrm{eV}$).
 
\textit{Discussion and summary -}
 In cuprates,  $\dx$ electrons act concurrently  as both carrier and strongly correlated medium where ``pairing glue" emerges~\cite{maier08}.  The entanglement of 
 the two roles of electrons may be one of the reasons why cuprate SC is so difficult to understand.  In pressurized $\lno$, the  two active $e_g$ orbitals may can  be  assigned separately with those two different roles for SC. This sheds new light on comprehending unconventional superconductivity from a novel perspective. 
 Future experiments and further theoretical studies on the BLTO-Hubbard model are warranted to confirm whether  the two-component theory~\cite{yang23inter,berg08,kivelson2002making} provides a substantial  description  of the SC
 in $\lno$. To summarize, in this work we study superconductivity in the BLTO-Hubbard model by using cluster DMFT methods. We 
 establish phase diagrams revealing the crucial dependence of superconducting $T_c$ on two key physical parameters,  $V$ and $t_{\perp}$ of the system. The relevance of the two-component theory of superconductivity to our results is discussed.

\textit{Acknowledgment- } We thank Meng Wang, A. -M. Tremblay, Dao-Xin Yao,  Changming Yue, and Xunwu Hu for useful discussions. W.W is in debt to Guang-Ming Zhang and Yi-feng Yang for discussions and suggestions. Work at Sun Yat-Sen University was supported by the National Natural Science Foundation of China (Grants  No.12274472, No.11904413).To reproduce the benchmark data of this work, one can employ the open source  DCA++ code (https://github.com/CompFUSE/DCA), with using necessary extension codes and data from ({https://github.com/WUWei20/BLTO\_CDMFT\_benchmark}).

\bibliography{Ni}

\end{document}